Report of the ESO Workshop

# Ground-based Thermal Infrared Astronomy – Past, Present and Future

held on-line, 12−16 October 2020

Leo Burtscher[1]
Valentin D. Ivanov[2]
Mario van den Ancker[2]

[1]Leiden Observatory, The Netherlands
[2]ESO

The ESO workshop *Ground-based thermal infrared astronomy* was held on-line October 12–16, 2020. Originally planned as a traditional in-person meeting at ESO in Garching in April 2020, it was rescheduled and transformed into a fully on-line event due to the COVID-19 pandemic. With 337 participants from 36 countries the workshop was a resounding success, demonstrating the wide interest of the astronomical community in the science goals and the toolkit of ground-based thermal infrared astronomy.

## Motivation

Observations in the thermal infrared (IR) regime (3–30 μm) provide a powerful tool to discover and characterise warm environments in the Universe from protoplanetary disks - the building sites of planets, to active galactic nuclei - the surroundings of accreting supermassive black holes. The thermal IR is also the wavelength range of choice to peek through exoatmospheric clouds, to characterise exoplanet atmospheres. Although space-based instruments offer the ultimate sensitivity, observations from the ground provide unrivalled spatial and spectral resolution. Thanks to regular upgrades, they are also the preferred testbed for new technologies or exciting experiments - as recently demonstrated by the NEAR experiment at the VLT (Kasper et al. 2019).

Astronomers working in the field routinely push instruments to their limits, demanding better sensitivity, stability on longer time scales and higher contrast which ultimately rely on complete instrument characterisation and calibration. This is not only very relevant for all major astronomical observatories which currently host thermal-IR cameras or spectrographs, such as VLT/VISIR, VLTI/MATISSE, and GranTeCan/CanariCam. Calibration in the thermal IR will be an even more important prerequisite for reaching the ambitious science goals of the next-generation facilities, e.g. characterising exoplanets is one of the prime science cases of METIS, a first-generation instrument for the Extremely Large Telescope (ELT) and of MICHI on the Thirty Meter Telescope (TMT).

The workshop brought together the experts in these fields to carry out two complementary types of review talks. Some of the speakers summarized the state of the art in individual fields with heavy usage of thermal-IR, such as solar system planets, studies of young stellar objects and evolved stars, the centre of the Milky Way galaxy and more distant active and star forming galaxies. Other speakers presented individual facilities, described their capabilities and paraded the most successful science cases addressed with these instruments. There was also a talk about the history of the field and a diversity and inclusion session. Last but not least, three discussion sessions on topics selected by the participants allowed for a lively exchange of opinions.

## Lessons from the On-line Workshop Format

The workshop was hosted online due to the ongoing COVID-19 pandemic, but also to increase inclusivity and sustainability. Three platforms facilitated the communication between the participants during the workshop (Fig. 1).

Talks and discussion sessions were transmitted live via Zoom. Slack was used to exchange text messages and to post files (e.g., PDFs of posters). Gather.town allowed video, audio and text interaction during the virtual coffee breaks, the virtual welcome reception and the dedicated poster viewing session. It was most popular during the poster session, while during the breaks the participants preferred to spend time offline. After the conference, two platforms were – and are – used for distributing content: video recordings of the presentations were placed on YouTube within 24 hours, to allow participants in different time zones to both watch them, and to participate in the discussions. Most of these talks remain publicly available on our YouTube channel[2]. Slides and posters were uploaded to Zenodo[3] to create a permanent record of the workshop that is also indexed by the SAO/NASA service ADS.

The selected workshop format was received very well by the conference attendees. The on-line discussion sessions were especially lively, with more than a thousand messages exchanged daily between the participants via Slack. However, having participants in many different time-zones proved to be a challenge. Some participants from e.g. the United States or Chile joined the live Zoom sessions in what was for them the middle of the night, others preferred to watch the talk recordings at a later time. While they were unable to fully participate in live discussions, Slack helped to follow-up on discussions the next day. For future on-line workshops an alternative format could be considered where all talks are pre-recorded and the live events are mostly limited to questions and discussions – but it remains to be seen if such a format leads to the same high level of interaction for the live participants as our fully live online event.

Last but not least, the online format very significantly reduced the carbon footprint of the workshop. If all 337 participants had traveled to Garching, an equivalent of 564 tons of $CO_2$ would have been emitted. Online conferences - and this one is no exception - are known to reduce emissions by at least a factor of three thousand compared to traditional real-life meetings (Burtscher et al. 2020).

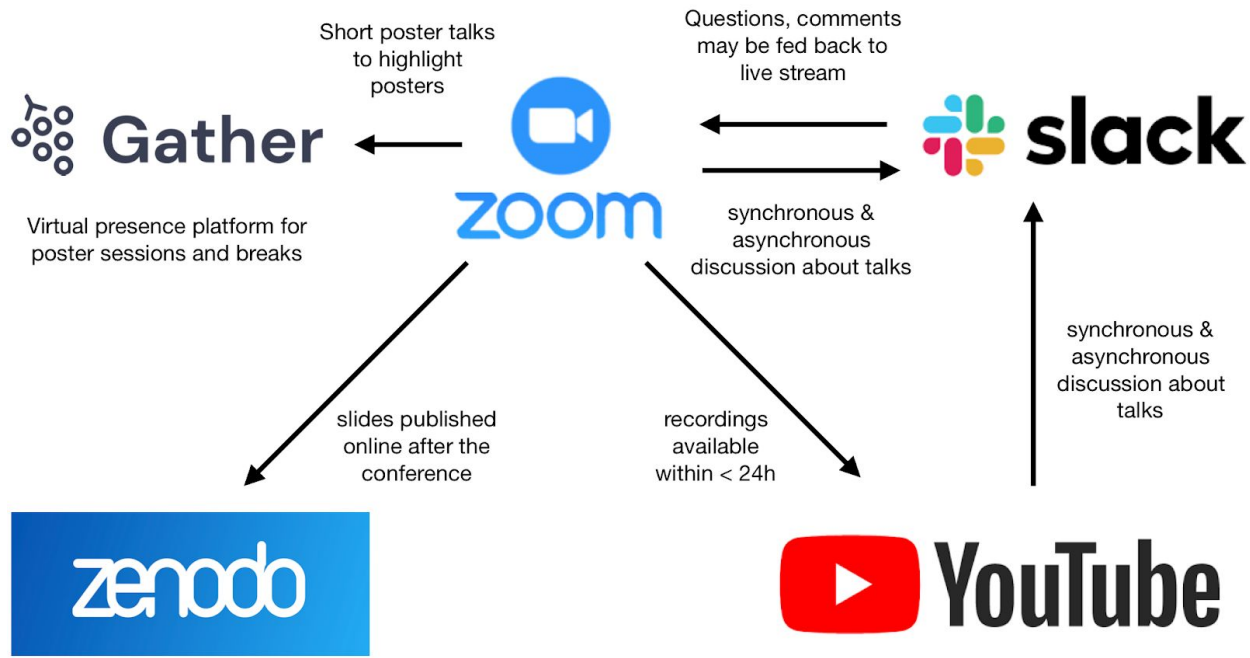

**Figure 1**. Platforms and services used to organize the on-line conference.

## Demographics and Inclusion

The workshop was well-attended, with 337 registered participants originating from 36 countries. Up to 160 people have been online at the same time. The typical number of participants for in-person ESO workshops during the recent years was 80-120 people. ESO member states dominated the demographics with 189 participants (56%), All ESO member states were represented. Women constituted 33% of the attendees, and the Scientific Organizing Committee formulated a scientific programme that reflected this percentage for the contributed talks (35 % female speakers). A special effort was made to maximize the fraction of **invited** female speakers, reaching a fraction of 39%. The selection of contributed talks was made without considering the gender or country of origin of the applicants. In addition, with 31% students, 23% postdocs and 46% senior researchers, the workshop achieved a good balance of career level and seniority - in fact, it was among our goals to promote the thermal-IR field among early-career astronomers.

Diversity and inclusion issues in astronomy were addressed specifically during a podium discussion led by Angela Speck in which a number of good practices were recommended. It is often already a good start to simply begin paying more attention to issues of diversity and inclusivity, use inclusive language, and be a "good bystander", e.g. intervene when microaggressions happen. The gender balance in astronomy was discussed, as well as specific issues such as the effect of the COVID-19 crisis on gender diversity. In addition, attention was paid to geographic and language barriers, and access of developing nations to the (expensive) infrastructure required to do observational astronomy in the thermal infrared.

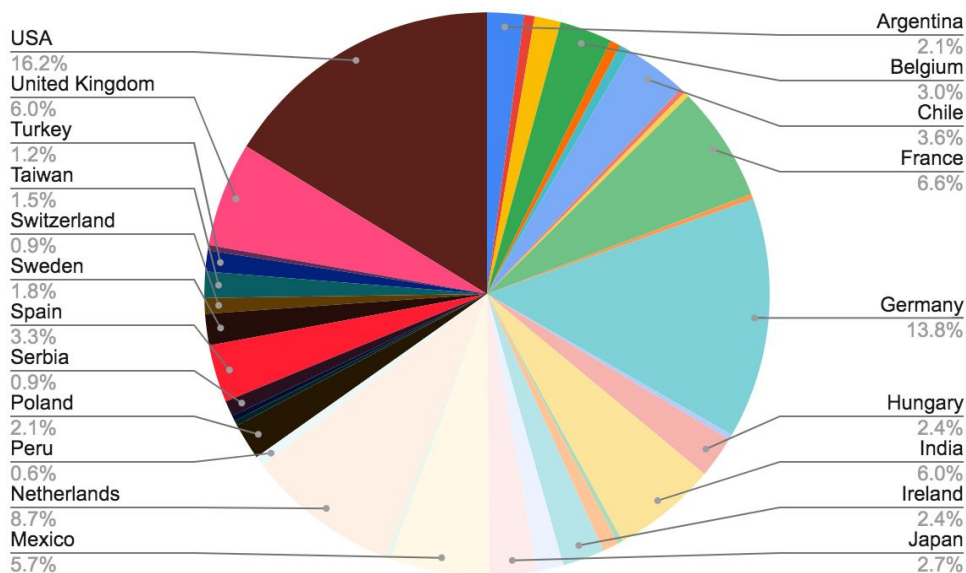

**Figure 2a**. Geographic distribution of workshop attendees.

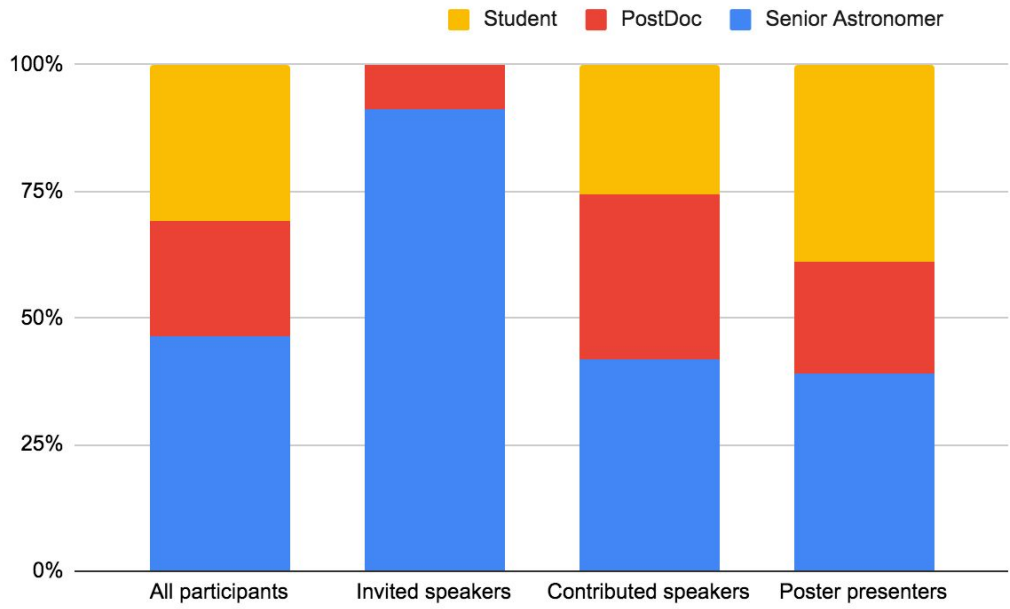

**Figure 2b**. Seniority level of workshop participants and presenters.

## Science Highlights

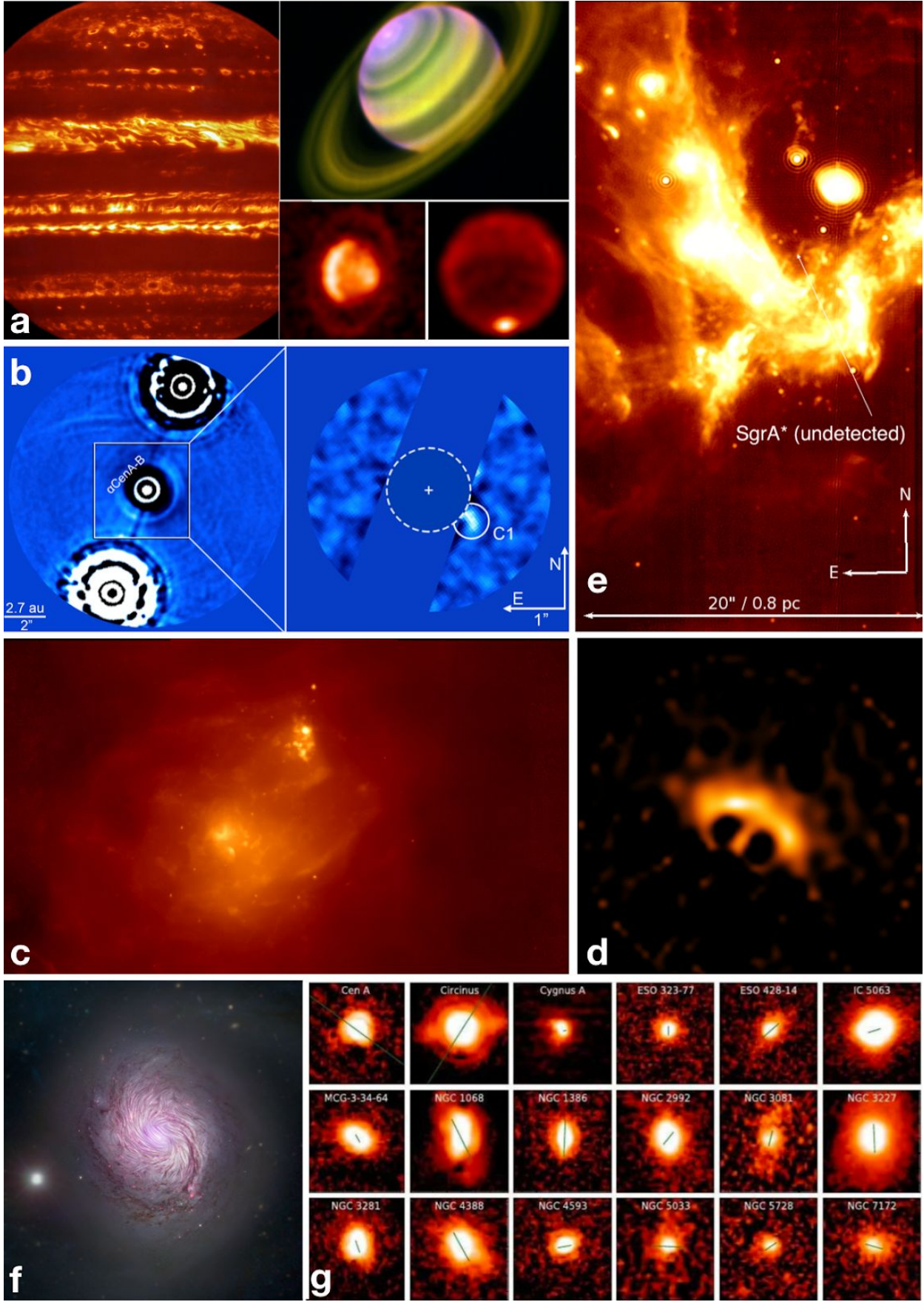

**Figure 3**. Visual and scientific highlights from ground-based thermal-IR observations at large facilities, ordered by distance from the Earth. See text for description.

Over the five days of the workshop a total of 63 talks were given (23 invited, 43 contributed) and 18 on-line posters were presented, divided into ten scientific areas (AGNs, Dust/ISM, Exoplanets, Galactic Center, Galaxies, Instruments, Protoplanetary Disks, Solar System, Stars / circumstellar environments and YSOs). In this short article, we cannot give a comprehensive review of all these fields, but based on a number of both scientifically and visually striking examples, we would like to highlight the diversity of science fields to which ground-based thermal-IR astronomy is contributing and how this observing technique provides a unique angle to these fields (Fig. 3):

From near to far, these contributions include:
  a) observations of the giant planets from VISIR: Jupiter, Saturn, Uranus, and Neptune (left to right; top to bottom; Fletcher et al. 2017, 2018; Roman et al. 2020; Sinclair et al. 2020). The observations use wider waveband coverage, denser temporal sampling, and more modern technology than available on space probes.
  b) VLT/NEAR image of the Alpha Centauri A/B system as observed behind a coronagraphic Vortex mask. No planet has been detected in this image, but a detection would have been possible down to ~ 400 µJy, corresponding to a contrast of $3.2 \times 10^{-6}$ compared to Alpha Cen. A candidate (C1) has been detected that requires follow-up observations for confirmation (Wagner et al. 2021). Ground-based AO coronagraphic imaging in the thermal-IR has a unique potential to detect faint Earth-like planets, at more extreme contrasts than possible from space.
  c) A wide-field (~ 5' x 3.5') mosaic of the Orion nebula in the mid-infrared showing the Trapezium region bottom left of the center and the BN/KL complex just North of the center of the image, as well as significant filamentary structure in between. Due to the requirement to remove the thermal background in a crowded field, the processing of these observations was technically challenging and to the best of our knowledge, it represents the widest angle observation taken in the thermal-IR from the ground with a large telescope so far, i.e. it is wide-angle **and** high spatial resolution (Robberto et al. 2005).
  d) Reconstructed image of FS CMa as observed with VLTI/MATISSE: The L-band aperture-synthesis image shows the inclined disk of the unclassified B[e] star FS CMa. One can see the central star and the bright, inner edge of the disk with an angular resolution of about 4 mas (FOV is 60x60 mas, wavelength range of 3.4–3.8 mu). The north-western disk rim is brighter than the south-eastern one as one is directly looking at the northwestern, inner disk rim wall. The inner dust-depleted hole has a size of about 6x12 mas (Hofmann et al. 2021, in prep.). The spatial resolution afforded by VLTI/MATISSE is unparalleled by other thermal-IR facilities. Only in the thermal-IR is the bulk of the disk seen.
  e) Deepest image of the Galactic Center at 8.6 µm: ca. 2 hour on-target exposure with VISIR in the PAH1 filter, July 2017. Reduction with speckle holography for maximum spatial resolution. Sgr A* itself is undetected since it is confused with the "Sgr A* ridge" at these wavelengths. Schödel et al. (in prep)
  f) SOFIA/HAWC+ magnetic field lines overlaid on a visible (HST, SDSS) and X-ray composite image of the nearby prototypical AGN NGC 1068. The image supports the "density wave theory" on how the spiral arms are forced into their iconic shape (Lopez-Rodriguez et al. 2020).
  g) Observations of a sample of mid-IR bright active galaxies showing that the central, AGN-heated dust continuum emission is elongated along the same direction as the polar axis of the AGN indicated by a green bar that is 100 pc in length. The AGN heated dust can only be resolved in thermal-IR observations from the ground (Asmus, Hönig & Gandhi 2016; Asmus 2019).

In addition, three discussion panels were held sequentially and with good and very lively participation from the community: In the first block, we asked what the different science areas can learn from each other, e.g. can we apply YSO disk+outflow models to AGNs or evolved stars? We found both similarities (e.g. how to link observables to models and which radiative transfer codes to use under which conditions) and differences (e.g. the apparent ubiquity of polar-oriented outflows in AGNs was contrasted with the disk-morphology of YSOs). A second discussion block was devoted to instrumentation development and addressed the question which instrumentation is required in order to pursue our research (and how to get it). In this session we discussed, among other suggestions, polarimetry and high spectral resolution in the N band (8–13 micron) on ELTs. A third discussion block was devoted to community building and investigated what support early career researchers require to enter the field of thermal-IR astronomy in particular, whether they need more broad conferences (like this one), a new text book or more extensive user support (e.g. like the ALMA support nodes). There was agreement among the participants that a follow-up conference in 1-2 years' time would be useful.

## Public Outreach Event and Social Media

To further the workshop's goal of promoting the awareness of thermal infrared astronomy, a public outreach event[4] was held as part of the workshop in collaboration with the "Haus der Astronomie" in Heidelberg. Nine public talks, given in seven different languages, were streamed live via the Haus der Astronomie's YouTube channel, reaching more than 2500 viewers. The videos are still accessible in their dedicated YouTube playlist[5]. In addition, highlights from the workshop were communicated via the @ESO_IR2020 channel on Twitter, with 155 tweets generating a further 18.900 impressions.

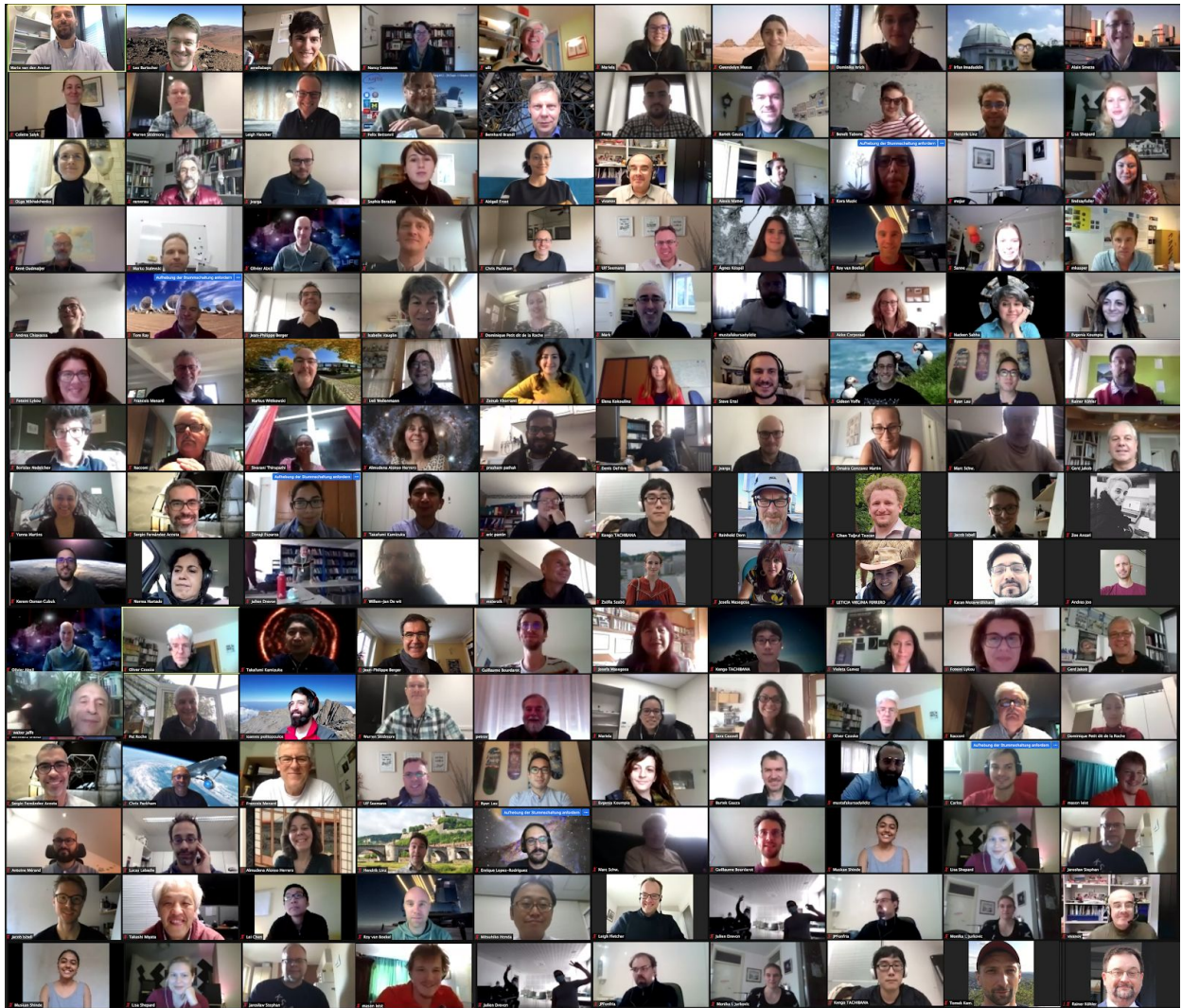

**Figure 4**. Virtual conference picture showing some of the participants following the live talks.

## Future Outlook

Thermal-infrared astronomy is entering a new era, both on the ground and in space, with the arrival of a number of new facilities. JWST is expected to launch later this year. It will provide access to a wavelength range up to 29 μm and will maximise sensitivity thanks to the low thermal background achievable in space. METIS at the ELT is currently expected to see first light in 2028 and will cover the range from 3 to 13 microns, delivering both diffraction limited imaging at the ELT resolution of 23 milli-arcseconds at 3.5 μm and spectroscopy with resolving powers from a few hundreds to a hundred thousand.

A number of innovative smaller instrument projects are also progressing well. For example, MIMIZUKU at the Tokyo Atacama Observatory (TAO) 6.5m telescope at 5640m altitude will reach the ultimate (ground-based) transmission, particularly in the challenging Q band (~ 20–30 μm). MIRAC-5 is planned to be upgraded with the novel HgCdTe-based "GeoSnap" detector by Teledyne, possibly a breakthrough in thermal-infrared detector technology. It will be operated both on the MMT and at the Magellan telescope – both with adaptive optics support – and will pave the way for the use of this novel detector technology in METIS at the ELT and MICHI at the TMT.

Nancy Levenson, in her invited review talk, made a point that the space and ground based facilities complement each other in sensitivity, collecting area, cost and rate of technological innovation. She underlined that big facilities serve as natural loci to form active user communities that use the telescopes, but also contribute via various mechanisms - decadal surveys, user committees, etc. - to their planning, development and operations. Our workshop demonstrated the existence of an vibrant thermal IR community. The meeting will help to ensure its growth and the inclusion of young astronomers, and to foster close ties amongst the community members for the future.

### Acknowledgments

The organizers would like to thank ESO and Leiden Observatory for sponsoring the workshop. We would also like to thank Nelma Silva, Véronique Ziegler, Jutta Boxheimer and Josh Carr of Three Counties Media without whose administrative support, help in graphic design of the poster and video editing it would not have been possible to organize the workshop.

### Links

[1] Link to workshop programme: https://www.eso.org/sci/meetings/2020/IR2020/program.html
[2] YouTube channel with recordings of most talks:
https://www.youtube.com/channel/UCsTNXi_Sa1j8HQaJAtpmLjg/playlists
[3] Presentations archived at Zenodo: https://zenodo.org/communities/esoir2020
[4] Public Outreach Event: https://www.eso.org/sci/meetings/2020/IR2020/public_talks.html
[5] YouTube playlist for the nine public talks organised for the IR 2020 workshop:
https://www.youtube.com/playlist?list=PL6v1Ej3QgEXU6L0cuIH0Imu-_8vpcTSB2